\documentclass{PoS}

\usepackage{epsf}
\usepackage{amssymb}
\usepackage{amsmath}
\usepackage{amsfonts}
\usepackage{cite}
\usepackage[small]{caption2}

\usepackage[T1]{fontenc}
\usepackage{psfrag,epsfig,graphicx,graphics}

\newcommand{\be}{\begin{equation}}
\newcommand{\beq}{\begin{equation}}
\newcommand{\ee}{\end{equation}}
\newcommand{\eq}{\end{equation}}
\newcommand{\eeq}{\end{equation}}

\newcommand{\bea}{\begin{eqnarray}}
\newcommand{\eea}{\end{eqnarray}}

\def\slashchar#1{\setbox0=\hbox{$#1$}
   \dimen0=\wd0
   \setbox1=\hbox{/} \dimen1=\wd1
   \ifdim\dimen0>\dimen1
      \rlap{\hbox to \dimen0{\hfil/\hfil}}
      #1
   \else
      \rlap{\hbox to \dimen1{\hfil$#1$\hfil}}
      /gdatdafinal2.tex
   \fi}

\newcommand{\fV}{f_{3\,\rho}^V}
\newcommand{\fA}{f_{3\,\rho}^A}

\title{Exclusive electroproduction of $ \rho_T$ meson with twist three accuracy}

\ShortTitle{Exclusive electroproduction with twist three accuracy}

\author{I. V. Anikin\\
        BLTP, JINR, 141980 Dubna, Russia\\
        E-mail: \email{anikin@theor.jinr.ru}
}

\author{D. Yu. Ivanov\\
        Sobolev Institute of Mathematics, 630090 Novosibirsk, Russia\\
        E-mail: \email{d-ivanov@math.nsc.ru}
}

\author{B. Pire\\
        CPhT, Ecole Polytechnique, CNRS, 91128 Palaiseau, France\\
       E-mail: \email{pire@cpht.polytechnique.fr}
}

\author{\speaker{L. Szymanowski}
\\
        SINS, Warsaw, Poland\\
        E-mail: \email{lechszym@fuw.edu.pl}}

\author{S.~Wallon
\\
LPT, Universit{\'e} Paris-Sud, CNRS, 91405 Orsay, France \ {\em \&} \\
UPMC Univ. Paris 06, facult\'e de physique, 4 place Jussieu, 75252 Paris Cedex
05, France\\
        E-mail: \email{wallon@th.u-psud.fr}}

\abstract{
Exclusive electroproduction of vector mesons is a crucial way to probe
QCD factorization properties. Such a factorization is up to now only
proven, at the twist 2 level, for a longitudinaly polarized meson. It is
crucial to extend our understanding to the case of transversely polarized
vector mesons. As a first step in this direction, we evaluate the
impact factor of the transition $\gamma^* \to \rho_T $, which is the
relevant part of the amplitude within the $k_t$-factorization approach valid
at large energies, taking into account the twist 3 contributions, coming
both from quark antiquark and from quark antiquark gluon correlators. We
show that a gauge invariant expression is obtained with the help of QCD
equations of motion.
}

\FullConference{35th International Conference of High Energy Physics - ICHEP2010,\\
		July 22-28, 2010\\
		Paris France}

\begin{document}

\section{Introduction}
\label{Sec_Int}

 The factorization properties of the leading twist amplitudes allow to study deep exclusive meson electroproduction \cite{fact}, when the meson is (pseudo)scalar or longitudinally polarized.
The case  of a transversely polarized vector meson is more involved since the chiral-oddity of its  leading twist distribution amplitude (DA)
leads to its decoupling in leading twist electroproduction amplitudes
 \cite{DGP} unless in reactions with more than two final
hadrons \cite{IPST}.
To understand available data  \cite{exp}, one thus needs to consider carefully all
twist 3 contributions.
The meson quark gluon structure within collinear factorization may be described by Distribution Amplitudes (DAs),
classified in \cite{BB}. We  consider here the case of  very high energy collisions at electron proton or electron positron colliders \cite{IP,PSW}.
In the literature there are two approaches to the factorization of the
scattering amplitudes in exclusive processes at leading and higher twists. 
The Light-Cone Collinear Factorization (LCCF) \cite{APT,AT} extends the inclusive approach \cite{EFP}
 to exclusive processes, dealing with the factorization in the momentum space around the dominant light-cone
direction, while  the Covariant Collinear Factorization (CCF) approach
in coordinate space was succesfully applied in \cite{BB} for a systematic
description of DAs of hadrons carrying different
twists.  
We show  \cite{us} that these two descriptions are equivalent at twist 3. 
For that, we perform
our analysis  within LCCF method in momentum space.
It introduces relevant soft correlators which are generally not independent ones.
The correlators are reduced  to a minimal  independent set  with the use of equation of motions and of the light-cone-fixing vector independence condition.
  A dictionary is obtained between LCCF and CCF correlators, proving the equivalence between LCCF and CCF approaches.
We  illustrate this equivalence by   calculating up to twist 3
accuracy
within both methods the impact factor $\gamma^* \to \rho_T$, 
which enters the description of the $\gamma^* \, p \to \rho \, p$  and 
$\gamma^*\, \gamma \,\to \,\rho \,\rho$ processes at large $s$. 

\section{LCCF factorization of exclusive processes}
 \label{Sec_LCCF}


The amplitude for the exclusive process $A \to \rho \, B$ is,  in
 the momentum representation and in axial
gauge reads ($H$ and $H_\mu$ are  2- and 3-parton coefficient functions,
 respectively)
\begin{eqnarray}
\label{GenAmp}
{\cal A}=
\int d^4\ell \, {\rm tr} \biggl[ H(\ell) \, \Phi (\ell) \biggr]+
\int d^4\ell_1\, d^4\ell_2\, {\rm tr}\biggl[
H_\mu(\ell_1, \ell_2) \, \Phi^{\mu} (\ell_1, \ell_2) \biggr] + \ldots \,.
\end{eqnarray}
In (\ref{GenAmp}), the soft parts $\Phi$ are  the
Fourier-transformed 2- or 3-parton correlators which are matrix elements of non-local operators.
To factorize the amplitude, we  choose the dominant direction around which
we  decompose our relevant momenta and  we Taylor expand the hard part.
Let $p\sim p_\rho$ and $n$ be two light-cone vectors ($p \cdot n =1$).  Any vector $\ell$ is then expanded as
\begin{eqnarray}
\label{k}
\ell_{i\, \mu} = y_i\,p_\mu  + (\ell_i\cdot p)\, n_\mu + \ell^\perp_{i\,\mu} ,
\quad y_i=\ell_i\cdot n ,
\end{eqnarray}
and  the integration measure in (\ref{GenAmp}) is replaced as
$d^4 \ell_i \longrightarrow d^4 \ell_i \, dy_i \, \delta(y_i-\ell\cdot n) .$
The hard part  $H(\ell)$ is then expanded around
the dominant  $p$ direction:
\begin{eqnarray}
\label{expand}
H(\ell) = H(y p) + \frac{\partial H(\ell)}{\partial \ell_\alpha} \biggl|_{\ell=y p}\biggr. \,
(\ell-y\,p)_\alpha + \ldots
\end{eqnarray}
where $(\ell-y\,p)_\alpha \approx \ell^\perp_\alpha$ up to twist 3.
To obtain a factorized amplitude, one performs 
an
 integration 
 by parts
to replace  $\ell^\perp_\alpha$ by $\partial^\perp_\alpha$ acting on
the soft correlator.
 This leads to new operators containing
transverse derivatives, such as $\bar \psi \, \partial^\perp \psi $,
 thus requiring
additional DAs
$\Phi^\perp (l)$.
Factorization is then achieved by 
 Fierz decomposition on a set of relevant Dirac $\Gamma$ matrices, and we end up with
\bea
\label{GenAmpFac23}
\hspace{-.4cm}{\cal A}=
  {\rm tr} \left[ H_{q \bar{q}}(y) \, \Gamma \right] \otimes \Phi_{q \bar{q}}^{\Gamma} (y)
+
 {\rm tr} \left[ H^{\perp\mu}_{q \bar{q}}(y)  \Gamma \right] \otimes \Phi^{\perp\Gamma}_{{q \bar{q}}\,\mu} (y) + {\rm tr} \left[ H_{q \bar{q}g}^\mu(y_1,y_2) \, \Gamma \right] \otimes \Phi^{\Gamma}_{{q \bar{q}g}\,\mu} (y_1,y_2) \,,
\eea
%
%
where $\otimes$ is the $y$-integration.
Although the fields coordinates $z_i$ are on the light-cone in both LCCF and CCF parametrizations of the soft non-local correlators,
 $z_i$ is along $n$ in LCCF while arbitrary in CCF.
The transverse physical polarization of the $\rho-$meson is defined by the conditions
\beq
\label{pol_RhoTdef}
e_T \cdot n=e_T \cdot p=0\,.
\eq
Keeping all the terms up to the twist-$3$ order
with the axial (light-like) gauge, $n \cdot A=0$,
the matrix elements of quark-antiquark nonlocal operators
 for vector and axial-vector correlators without and with transverse derivatives,
with $\stackrel{\longleftrightarrow}
{\partial_{\rho}}=\frac{1}{2}(\stackrel{\longrightarrow}
{\partial_{\rho}}-\stackrel{\longleftarrow}{\partial_{\rho}})\,,$
can be written 
as (here, $z=\lambda n$)
\begin{eqnarray}
\label{par1v}
\langle \rho(p_\rho)|\bar\psi(z)\gamma_{\mu} \psi(0)|0\rangle
&=&
m_\rho\,f_\rho \int_{0}^{1}\, dy \, {\rm exp}\left[iy\,p\cdot z\right] \left[ \varphi_1(y)\, (e^*\cdot n)p_{\mu}+\varphi_3(y)\, e^*_{T\mu}\right],\,\,\,\,
\\
\label{par1.1v}
\langle \rho(p_\rho)|
\bar\psi(z)\gamma_{\mu}
i\stackrel{\longleftrightarrow}
{\partial^T_{\alpha}} \psi(0)|0 \rangle
&=& m_\rho\,f_\rho \,\int_{0}^{1}\, dy \, {\rm exp}\left[iy\,p\cdot z\right]
\varphi_1^T(y) \, p_{\mu} e^*_{T\alpha}\,, \\
\label{par1a}
\langle \rho(p_\rho)|
\bar\psi(z)\gamma_5\gamma_{\mu} \psi(0) |0\rangle &=&
m_\rho\,f_\rho \, i\int_{0}^{1}\, dy \, {\rm exp}\left[iy\,p\cdot z\right]\varphi_A(y)\, \varepsilon_{\mu\alpha\beta\delta}\,
e^{*\alpha}_{T}p^{\beta}n^{\delta} \,,  \\
\label{par1.1a}
\langle \rho(p_\rho)| \bar\psi(z)\gamma_5\gamma_{\mu}
i\stackrel{\longleftrightarrow}
{\partial^T_{\alpha}} \psi(0) |0\rangle &=&
m_\rho\,f_\rho \,
i\int_{0}^{1}\, dy \, {\rm exp}\left[iy\,p\cdot z\right]\varphi_A^T (y) \, p_{\mu}\, \varepsilon_{\alpha\lambda\beta\delta}\,
e_T^{*\lambda} p^{\beta}\,n^{\delta}\,,
\end{eqnarray}
where 
$y$ ($\bar y$) is the quark (antiquark) momentum fraction.
Two analogous correlators are needed to describe gluonic degrees of freedom, introducing $B$ and $D$ DAs according to
\begin{eqnarray}
\label{Correlator3BodyV}
\hspace{-.8cm}\langle \rho(p_\rho)|
\bar\psi(z_1)\gamma_{\mu}g A_{\alpha}^T(z_2) \psi(0) |0\rangle \!
&=&\! m_\rho \,\fV \!
\int\limits_{0}^{1} \! dy_1 \! \int\limits_{0}^{1} \! dy_2 \,
e^{iy_1\,p\cdot z_1+i(y_2-y_1)\,p\cdot z_2 } \,
B(y_1,y_2) p_{\mu} e^*_{T\alpha}\,, \ \
\\
\label{Correlator3BodyA}
\hspace{-.6cm}\langle \rho(p_\rho)|
\bar\psi(z_1)\gamma_5\gamma_{\mu} g A_{\alpha}^T(z_2) \psi(0) |0\rangle
&=&m_\rho\,\fA \,
\int\limits_{0}^{1} dy_1 \,\int\limits_{0}^{1} dy_2 \,
e^{iy_1\,p\cdot z_1+i(y_2-y_1)\,p\cdot z_2 } \,
i D(y_1,y_2) \nonumber\\
&&\hspace{1cm}\times p_{\mu} \, \varepsilon_{\alpha\lambda\beta\delta} \,
e^{* \, \lambda}_T \, p^{\beta}n^{\delta}\,.
\end{eqnarray}
 One thus needs 7 DAs:  $\varphi_1$ (twist-$2$), 
$B$ and $D$ (genuine (dynamical) twist-$3$) and
$\varphi_3$, $\varphi_A, \varphi_1^T$, $\varphi_A^T$
(kinematical (\`a la
Wandzura-Wilczek) twist-$3$ and genuine (dynamical) twist-$3$).

These DAs are not independent.
They are related by 2 Equations of Motions (EOMs) and 2 equations arising from the invariance of  ${\cal A}$ under the arbitrary vector
$n,$
which  comes from 3 sources.
 First, it enters the definition
of the non-local correlators
 through
the light-like separation $z=\lambda \, n$.
%
These correlators are defined in the axial light-like gauge $n \cdot A=0\,,$
which allows to get rid of  Wilson lines.
Second, it determines the notion of transverse polarization of the $\rho\,.$
Last, $n$ enters the Sudakov decomposition
(\ref{k}) which defines the transverse parton momentum involved in the collinear factorization. One can in fact show that the hard part does not depend on the gauge fixing vector $n.$ Therefore, only the second and third source of $n-$dependence should be investigated.  Based on Ward identities, this $n-$dependence of 
${\cal A}$ can be recast 
in a system of constraints which only
involve  
the soft part.
We thus have only 3 independent DAs $\varphi_1$ , 
$B$ and $D$, which fully encode  \pagebreak

\noindent 
the non-perturbative content of the $\rho$ at twist 3.

 The original CCF parametrizations of the $\rho$ DAs~\cite{BB} also  involve 3 independent DAs, defined through 4  correlators related by EOMs. The 2-parton axial-vector correlator reads, 
\beq
\label{BBA}
\langle \rho(p_\rho)|\bar \psi(z) \, [z,\, 0] \, \gamma_\mu \gamma_5 \psi(0)|0\rangle =
\frac{1}{4}f_\rho\,m_\rho\, \varepsilon_\mu^{\,\,\,\alpha \beta \gamma} e^*_{T \alpha} \,p_\beta \, z_\gamma\,    \int\limits_0^1\,dy\,e^{iy(p \cdot z)}\,g_\perp^{(a)}(y)\;,
\eeq
%
$[z_1, \, z_2] = P \exp \left[ i g \int\limits^1_0 dt \, (z_1-z_2)_\mu A^\mu(t \,z_1 +(1-t)\,z_2)    \right]$
being the Wilson line. Denoting the meson polarization vector by $e,$
 $e_T$ is here defined to be orthogonal to the light-cone vectors $p$ and $z$:
\beq
\label{pol_Rho}
e_{T \mu}=e_\mu -p_\mu \frac{e \cdot z}{p \cdot z}-z_\mu \frac{e \cdot p}{p \cdot z} \, .
\eeq
Thus  $e_T$  (\ref{pol_Rho}) in CCF and $e_T$ 
(\ref{pol_RhoTdef})
 in LCCF differ since  $z$ does not generally point in the $n$ direction.
The  2-parton vector correlator reads (up to twist 3)
\beq
\label{BBV1}
\langle \rho(p_\rho)|\bar \psi(z) \, [z,\, 0] \, \gamma_\mu  \psi(0)|0\rangle = f_\rho\,m_\rho\int\limits_0^1\,dy\,e^{iy(p\cdot z)}\left[
p_\mu\,\frac{e^*\cdot z}{p\cdot z}\phi_{\parallel}(y) +
e^*_{T\mu}\,g_\perp^{(v)}(y) 
\right]
\,.
\eq
The 3-parton correlators are parametrized 
 (up to twist 3 level) according to
\bea
\hspace{-1cm}\langle \rho(p_\rho)|\bar \psi(z)[z,t\, z]\gamma_\alpha g \, G_{\mu\nu}(t\, z)[t\,z,0] \psi(0)|0 \rangle &=&
-i p_\alpha [p_\mu e^*_{\perp \nu}-p_\nu e^*_{\perp \mu} ] \, m_\rho \, \fV \nonumber \\
&\times& \int D \alpha \, V(\alpha_1,\alpha_2) \,
e^{\,i p \cdot z \,(\alpha_1+\,t\,\alpha_g)} \, , \label{GV}\\
\hspace{-1cm}\langle \rho(p_\rho)|\bar \psi(z)[z,t\, z]\gamma_\alpha\gamma_5 g \, \tilde G_{\mu\nu}(t\, z)[t\,z,0] \psi(0)|0 \rangle &=&
- p_\alpha [p_\mu e^*_{\perp \nu}-p_\nu e^*_{\perp \mu} ] \, m_\rho \,\fA \nonumber \\
&\times & \int D \alpha \, A(\alpha_1,\alpha_2) \,
e^{\,i \, p \cdot z \,(\alpha_1+\,t\,\alpha_g)} \,,\label{GA}
\eea
where $\alpha_1$, $\alpha_2$, $\alpha_g$ are momentum fractions of quark, antiquark and gluon respectively inside the $\rho-$meson, 
$\int D \alpha =\int\limits^1_0 d\alpha_1\int\limits^1_0 d\alpha_2 \int\limits^1_0 d\alpha_g\,
\delta(1-\alpha_1-\alpha_2-\alpha_g)$
and $\tilde G_{\mu\nu}=-{1\over 2}\epsilon_{\mu\nu\alpha\beta}G^{\alpha\beta}.$ 
A comparison of the  correlators (\ref{par1v}, \ref{par1.1v}, \ref{par1a}, \ref{par1.1a}, \ref{Correlator3BodyV}, \ref{Correlator3BodyA}) and (\ref{BBA}, \ref{BBV1}, \ref{GV}, \ref{GA})  in the axial gauge $n \cdot A=0$ gives the following
identification of the 2- and 3-parton DAs in LCCF and CCF  approaches:
\begin{eqnarray}
\label{relBBvector-axial}
&&\varphi_1(y)=
\phi_{\parallel}(y) ,
\quad
\varphi_3(y)=
 g_\perp^{(v)}(y) \,, \quad
\varphi_A(y) =
-\frac{1}{4} \, \frac{\partial g_\perp^{(a)}(y)}{\partial y}\,,
\\
\label{DictB-D}
 &&B(y_1,\,y_2)=-\frac{V(y_1, \, 1-y_2)}{y_2-y_1}\, \quad
D(y_1,\,y_2)=-\frac{A(y_1, \, 1-y_2)}{y_2-y_1}\,.
\end{eqnarray}



\section{$\gamma^* \to \rho_T$ Impact factor up to  twist three accuracy in LCCF and CCF}

We have calculated, in both LCCF and CCF, the forward impact factor $\Phi^{\gamma^*\to\rho}$ 
of the subprocess
 $g+\gamma^*\to g+\rho_T\,,$
 defined as
the integral of the  discontinuity in the $s$
channel of the  off-shell S-matrix element 
 ${\cal S}^{\gamma^*_T\, g\to\rho_T\, g}_\mu$.
  In LCCF, one computes the diagrams perturbatively in a fairly direct way,
which makes the use of the  CCF
 parametrization \cite{BB} less practical.
 We need to express the impact factor in terms of 
hard coefficient functions and soft parts parametrized by the light-cone  matrix
 elements. The standard technique 
here is an operator product expansion on the light cone, which
gives the leading term in the power counting.
 Since there is no operator definition for \linebreak 
an  impact
factor, we have to rely on perturbation theory. The primary complication encountered is
\pagebreak

\noindent
  that the $z^2\to 0$ limit of any single diagram is given in terms 
of  
light-cone 
matrix elements 
%
without any 
Wilson
line insertion between the quark and gluon operators (''perturbative correlators``), like
$
\langle \rho(p_\rho)|\bar \psi(z)\gamma_\mu \psi(0)|0 \rangle\,.$
Despite working in the axial gauge one cannot neglect  effects coming from the Wilson
lines since the  two light cone vectors $z$ and $n$ are not identical and thus, generically, Wilson lines are not equal to unity. Nevertheless in the axial gauge the contribution of each additional parton costs one extra power 
of $1/Q$, allowing the calculation to be
organized in a simple iterative manner expanding the Wilson line. 
At twist 3, we need to keep the contribution
$[z,0]=1+i \,g \int\limits^1_0 dt \, z^\alpha A_\alpha (z t)$ and to care about the difference between  the physical  $\rho_T$-polarization 
(\ref{pol_RhoTdef})
 from the formal one  (\ref{pol_Rho}).
At twist 3-level the net effect of the Wilson line when computing our impact factor is
just a renormalization of the DA  $g^a_\perp$  of  (\ref{BBA}), and similarly for the vector case.
We are then able to show that   our two LCCF and CCF results are identical;
the result is gauge invariant due to a consistent inclusion of fermionic and gluonic
degrees of freedom and  it is
 free of end-point singularities, due to the $k_T$
regulator.

This establishes a consistent gauge invariant analysis of electroproduction of
transversely polarized vector mesons at high energy. An extension of this work
to lower energy regime where collinear factorization allows to write the
amplitude in terms of generalized parton distributions is under way.

 This work is partly supported by  the grant
ANR-06-JCJC-0084, the RFBR (grants 09-02-01149,
09-02-00263,
08-02-00896), the grant 
NSh-3810.2010.2 
and
the Polish Grant N202 249235.

\end{document}